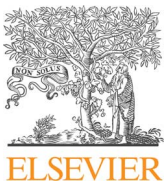
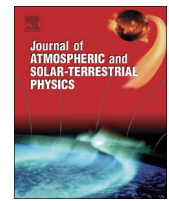
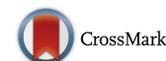

# Modelling the descent of nitric oxide during the elevated stratopause event of January 2013

Yvan J. Orsolini[a,b,*], Varavut Limpasuvan[c], Kristell Pérot[d], Patrick Espy[b,e], Robert Hibbins[b,e], Stefan Lossow[f], Katarina Raaholt Larsson[d], Donal Murtagh[d]

[a] *Norwegian Institute for Air Research (NILU), Instituttveien 18, N-2027, Kjeller, Norway*
[b] *Birkeland Centre for Space Science, University of Bergen, Postboks 7803, N-5020 Bergen, Norway*
[c] *School of the Coastal Environment, Coastal Carolina University, Conway, SC 29526, USA*
[d] *Department of Earth and Space Sciences, Chalmers University of Technology, SE-412 96 Gothenburg, Sweden*
[e] *Department of Physics, Norwegian University of Science and Technology (NTNU), NO-7491 Trondheim, Norway*
[f] *Karlsruhe Institute of Technology, IMK-ASF, P.O. Box 3640, 76021 Karlsruhe, Germany*

ABSTRACT

Using simulations with a whole-atmosphere chemistry-climate model nudged by meteorological analyses, global satellite observations of nitrogen oxide (NO) and water vapour by the Sub-Millimetre Radiometer instrument (SMR), of temperature by the Microwave Limb Sounder (MLS), as well as local radar observations, this study examines the recent major stratospheric sudden warming accompanied by an elevated stratopause event (ESE) that occurred in January 2013. We examine dynamical processes during the ESE, including the role of planetary wave, gravity wave and tidal forcing on the initiation of the descent in the mesosphere-lower thermosphere (MLT) and its continuation throughout the mesosphere and stratosphere, as well as the impact of model eddy diffusion. We analyse the transport of NO and find the model underestimates the large descent of NO compared to SMR observations. We demonstrate that the discrepancy arises abruptly in the MLT region at a time when the resolved wave forcing and the planetary wave activity increase, just before the elevated stratopause reforms. The discrepancy persists despite doubling the model eddy diffusion. While the simulations reproduce an enhancement of the semi-diurnal tide following the onset of the 2013 SSW, corroborating new meteor radar observations at high northern latitudes over Trondheim (63.4°N), the modelled tidal contribution to the forcing of the mean meridional circulation and to the descent is a small portion of the resolved wave forcing, and lags it by about ten days.

## 1. Introduction

The origin and lifecycle of stratospheric sudden warmings (SSWs) through the interaction of upward-propagating planetary waves with the zonal mean flow have been widely studied since the seminal paper of Matsuno (1971). Yet, the dramatic changes that occur in both the middle and the upper atmosphere during SSWs continue to receive attention, spurred by improved satellite and ground-based observations, whole-atmosphere model development and extension of global meteorological analyses into the mesosphere.

Manney et al. (2008, 2009) used observations from the NASA's Microwave Limb Sounder (MLS) instrument to show how strong SSWs were accompanied by the formation of a separated elevated polar stratopause. Several facets of the coupling between the stratosphere, mesosphere and lower thermosphere at high latitudes during such events, now referred to as Elevated Stratopause Events (ESEs), have since been studied in great detail using other observational datasets as well as models. These key facets are: (i) the polar cooling in the mesosphere that accompanies the stratospheric warming; (ii) the polar lower thermosphere warming aloft, first predicted in model studies (Liu and Roble, 2002); (iii) the reformation of the stratopause at standard mesospheric heights, and the subsequent mesospheric descent that occurs as the elevated stratopause returns to its climatological altitude (Siskind et al., 2010; Orsolini et al., 2010; Funke et al., 2010; Yamashita et al., 2010; Smith et al., 2011; Limpasuvan et al., 2012; Kvissel et al., 2012; Tomikawa et al., 2012; Chandran et al., 2013, 2014; Zülicke and Becker, 2013). In addition, the ESEs accompany changes at mid and low latitudes in the middle atmosphere (e.g. Kishore et al., 2016; Limpasuvan et al., 2016). They also alter the gravity wave-induced coupling of the lower atmosphere with the

---

thermosphere (e.g. Yiğit and Medvedev, 2016). Model studies further revealed that the reformation of the elevated polar stratopause is initially influenced by forcing from planetary waves, before the gravity wave drag regains its dominant role in forcing the downward branch of the mean meridional circulation (Limpasuvan et al., 2012, 2016). Enhancement of planetary wave activity in the mesosphere-lower thermosphere (MLT) during ESEs has been confirmed by radar observations over the years 2000–2008 (Stray et al., 2015). This mesospheric descent is particularly important for the transport of nitric oxides ($NO_x=NO+NO_2$) produced in the MLT by energetic particle precipitation (EPP). Since $NO_x$ is long-lived in polar night conditions and drives catalytic destruction of stratospheric ozone when the sun returns, its abundance can affect stratospheric temperatures and dynamics. Transport of sufficient $NO_x$ amount from its MLT reservoir into the stratosphere is central for a chemistry-climate model to represent the EPP effect on stratospheric composition and dynamics. Several recent case studies point out that chemistry-climate models are still deficient in that respect, hindering a full representation of the solar impact upon climate (Randall et al., 2015; Siskind et al., 2015).

A series of observational and model studies have also pointed out the enhancement of the semi-diurnal tide following ESEs. Bhattacharya et al. (2004) used ground-based measurements of winds near the mesopause above northern Canada (74°N) to show amplification of the semi-diurnal tide. More recently, Goncharenko et al. (2012) and Pedatella and Forbes (2010) observed the semi-diurnal tidal enhancement following the major SSWs in 2009 in total ionospheric electron content. Sridharan et al. (2012) and Nozawa et al. (2012) also reported the enhancement of that tidal component in case studies of SSWs using radar wind observations at low latitudes and high latitudes, respectively. Jin et al. (2012) also observed the enhancement of the migrating semi-diurnal tide [SW2] in the Sounding of the Atmosphere using Broadband Emission Radiometry (SABER) temperature at low and mid latitudes. Model case studies using the Whole Atmosphere-Ionosphere Couple Model (GAIA) (Liu et al., 2013), the extended Whole Atmosphere Community Climate Mode (WACCM-X) (Sassi et al., 2013) or the Whole Atmosphere Model (WAM) model (Wang et al., 2014), and a model composite study using the specified dynamics version of WACCM (SD-WACCM) (Limpasuvan et al., 2016) reproduced the SW2 tidal amplification following SSWs. These studies linked the amplification to either an increase in low-latitude ozone (due to cooling), which is a source of tidal forcing, or to a change in background winds and vorticity, which can modulate the tidal waveguide (McLandress, 2002).

However, the relative role of the planetary, gravity and tidal waves in driving the descent from the MLT is not fully understood. In addition to the aforementioned wave processes, eddy and molecular diffusion are also important for the transport of vertically stratified trace species in the MLT (Garcia et al., 2014; Meraner et al., 2016).

In this study, we examine the recent ESE that occurred in January 2013, and its impact of the NO (a key component of $NO_x$) descent into the stratosphere. The results are twofold. We first show that simulations with a whole-atmosphere climate model nudged by meteorological analyses still underestimate the large descent of NO compared to satellite observations from the Sub-Millimetre Radiometer instrument (SMR) instrument. This strong descent starting in the MLT region is also clearly detected in the water vapour observations from SMR. Secondly, we demonstrate that the discrepancy arises abruptly at a time when planetary wave activity increases, just before the elevated stratopause reforms. While the simulations reproduce an enhancement of the SW2 tide following the onset of the ESE and corroborates Meteor radar observations at Trondheim (63.4°N), the tidal contribution to the forcing of the mean meridional circulation is small in comparison to the planetary wave forcing.

## 2. Modelling approach and observations

### 2.1. Model SD-WACCM

We use the National Center for Atmospheric Research (NCAR) Whole Atmosphere Community Climate Model with specified dynamics (SD-WACCM) of the. WACCM is a global circulation model with fully coupled chemistry and dynamics, and extends from the surface to ~145 km (with 88 pressure levels in total). The specified dynamics version relaxes the model's dynamics and temperature up to about ~0.79 hPa toward the Modern-Era Retrospective Analysis for Research and Application (MERRA) reanalysis of NASA's Global Modelling and Assimilation Office (Rienecker et al., 2011). Above this level, SD-WACCM transitions linearly to a free running model, and is fully interactive above 0.19 hPa. While nudging constrains the model's troposphere and stratosphere to follow the reanalysis, the meteorological and tracer response of SD-WACCM above 0.79 hPa has been shown to represent reasonably well the observed MLT during SSW ESE events (Tweedy et al., 2013; Chandran et al., 2014; Limpasuvan et al., 2016).

Model simulations were started on October 1, 2012 from an existing SD-WACCM initial state. The model was run at a horizontal resolution of 1.9°×2.5° (latitude×longitude), and the simulated fields were output every 3 h. Spectral analysis of our model simulations with 3-hourly output allows estimation of the amplitude of the diurnal, semidiurnal and terdiurnal migrating or non-migrating tides. Following Tweedy et al. (2013), the reversal of the zonal-mean zonal wind from westerlies to easterlies at the 1 hPa level (~50 km) is defined as the onset, which occurred on January 5. We refer to this onset day as Day 0. The stratopause location is based on the local temperature maximum between 20 km and 100 km. The residual mean meridional circulation and eddy forcings are analysed in the framework of the transformed Eulerian mean (TEM) formalism and of Eliassen-Palm fluxes (EP flux) (Andrews et al., 1987).

Two SD-WACCM simulations have been performed with standard and enhanced eddy diffusion. Halving the equivalent Prandtl number ($Pr$) from the standard value in WACCM ($Pr=4$) to $Pr=2$ enhances the rate of eddy diffusion applied to heat and trace species (Smith, 2012). This adjustment has been shown to improve the mean vertical profile of species with a strong vertical gradient across the mesopause, such as carbon dioxide or monoxide, by allowing more gravity wave-induced mixing in the MLT region (Garcia et al., 2014).

### 2.2. Satellite observations

We use the global distribution of temperature from Level 2 retrievals of NASA MLS version 3.3/3.4 (Livesey et al., 2011). MLS profiles along the orbital tracks cover approximately 82°S to 82°N, and the vertical coverage extends from 15 km to 90 km. We also use measurements of NO and water vapour from SMR aboard the Odin satellite, a limb emission sounder measuring globally a variety of trace gases as well as temperature in the middle atmosphere. SMR observes NO thermal emission lines in a band centred near 551.7 GHz. The vertical resolution is 7 km in the upper stratosphere and in the mesosphere. In the period considered here, NO data are available on an irregular basis of two observational days in a 14-day cycle (approximately 4 days per month). Water vapour ($H_2O$) is retrieved from a strong thermal emission line at 557 GHz, providing information from 40 to 100 km with an altitude resolution of about 3 km. We use the latest improved version V2.3 here, and more background information on SMR retrievals can be found in Pérot et al. (2014) and Lossow et al. (2009). Finally, we also use temperature from SABER aboard the NASA's Thermosphere Ionosphere Mesosphere Energetics and Dynamics satellite (Remsberg et al., 2008). Available since 2002, the SABER profiles extend from altitudes of 20–110 km and continuously cover [52°S–52°N]. Latitudes poleward of 52° are sampled only on successive 60-day yaw periods.

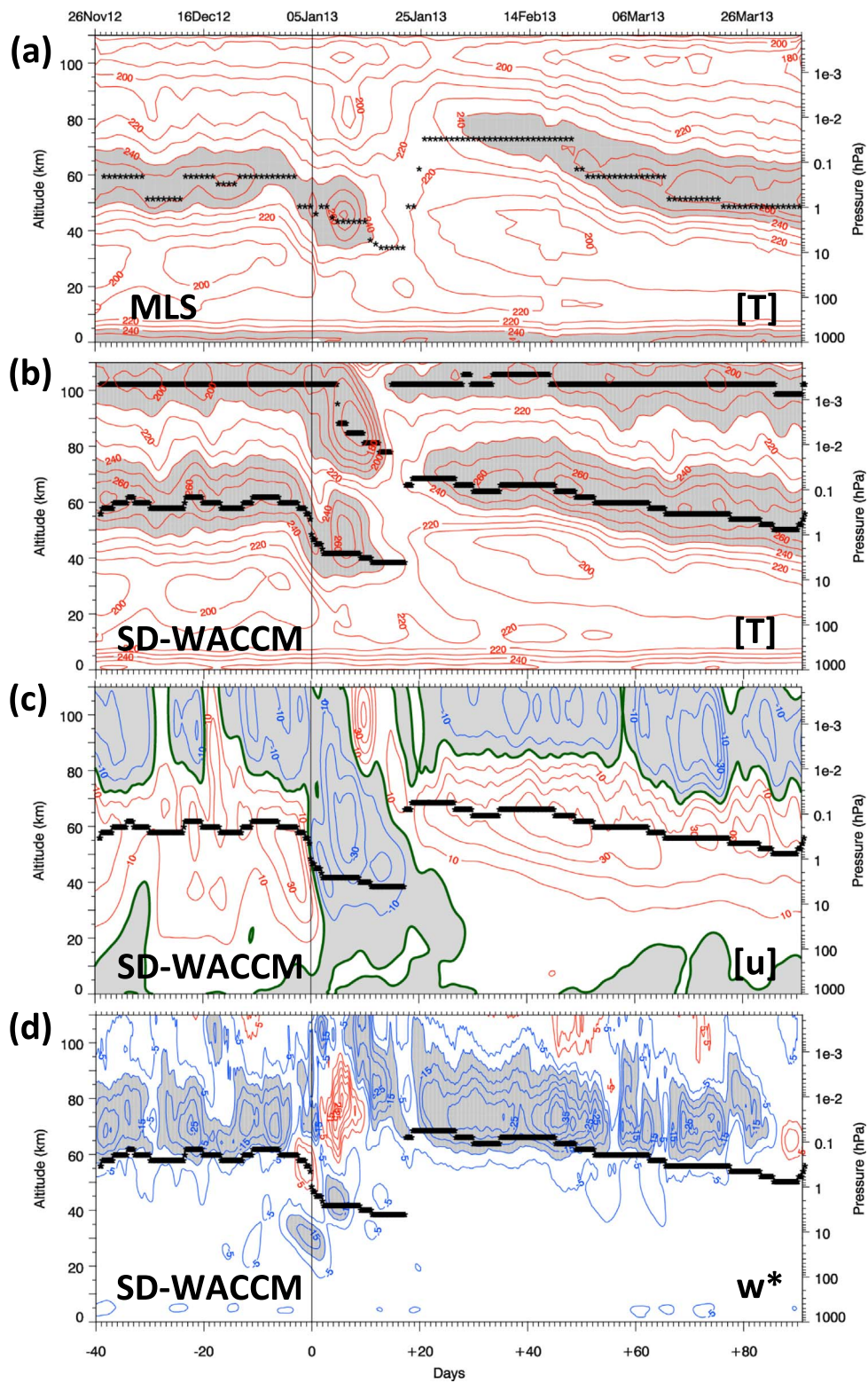

**Fig. 1.** Zonal-mean (A) temperature from the MLS observations (°K), (B) temperature (°K), (C) zonal wind (m s$^{-1}$), and (D) vertical TEM velocity (mm s$^{-1}$) from SD-WACCM. All quantities are averaged from 70°N to 90°N. The stars indicate the location of the stratopause below 80 km, and the mesopause above 80 km in (B). Days relative to the SSW onset are indicated as x-axis (bottom). Shaded areas correspond to temperatures higher than 240 K (A) and (B), to easterlies (C) or to velocities more negative than −10 mm s$^{-1}$.

### 2.3. Radar data

To characterise the semi-diurnal tide variability during the SSW, we use meteor wind observations from a SkiYMET all sky meteor radar located at Trondheim, Norway (63.4°N,10.5°E). Operating at 34.21 MHz with a pulse repetition frequency of 925 Hz, the transmitter consists of 8 three-element Yagi antennas in a circular array with a peak power of 30 kW. The return signal is received on a standard 5-antenna array. de Wit et al. (2015b) provides more detail on the radar system. Between 1 December 2012 and 1 March 2013, the radar ran without interruption, typically detecting around 10,000 unambiguous meteors per day between 75 and 100 km altitude. Hourly mean two component horizontal winds are fit to the position and line-of-sight velocities of the individual meteor trail detections in 6 separate altitude bins centred around 82, 85, 88, 91, 94 and 98 km. For each altitude bin, sine waves with periods 48, 24, 12 and 8 h (representing the quasi two-day wave and the diurnal, semidiurnal and terdiurnal tide, respectively) and an offset (representing the mean background wind) were fit each day to four-day running segments of the hourly mean horizontal winds.

### 3. Zonal-mean dynamical and thermal structure during the ESE of January 2013

We first examine polar-averaged [70–90°N] temperatures observed by the MLS instrument, between the surface and 110 km, during late November to early April (Fig. 1A). The vertical line denotes the SSW onset on January 5, while the stars indicate the approximate heights of the stratopause and mesopause. Shortly before the onset, the polar stratopause begins to descend and reaches down to near 35 km while, at the same time, the mesosphere cools and the temperature drops to 180 K near 105 km. This cooling appears to be a part of the downward propagating cold anomalies during SSW (Limpasuvan et al., 2016) and associated with the drop in the mesopause (e.g., Kishore et al., 2016). An elevated stratopause reforms near 75 km about 20 days after the onset, and remains at its maximum altitude until mid-February, before descending to its pre-warming altitude near 50 km over the following month. While the stratopause in SD-WACCM timely descends and reforms in good correspondence to MLS observations, its uppermost altitude is nevertheless about 5 km lower (Fig. 1B). The mesospheric cooling is also more pronounced in the SD-WACCM simulations, with the temperature reaching down to 170 K near 85 km.

Modelled zonal-mean zonal winds at high latitudes in Fig. 1C show that while easterlies prevailed in the pre-warming period only above 80 km, they rapidly appear into the stratosphere following the onset, where they persist longer. After the formation of the elevated stratopause, the stratospheric eastward jet recovers and gets stronger than in the pre-warming period, while the westward regime is re-established above 80 km after a 10-day interruption following the onset. Fig. 1D also shows the time evolution of the residual TEM vertical velocity.

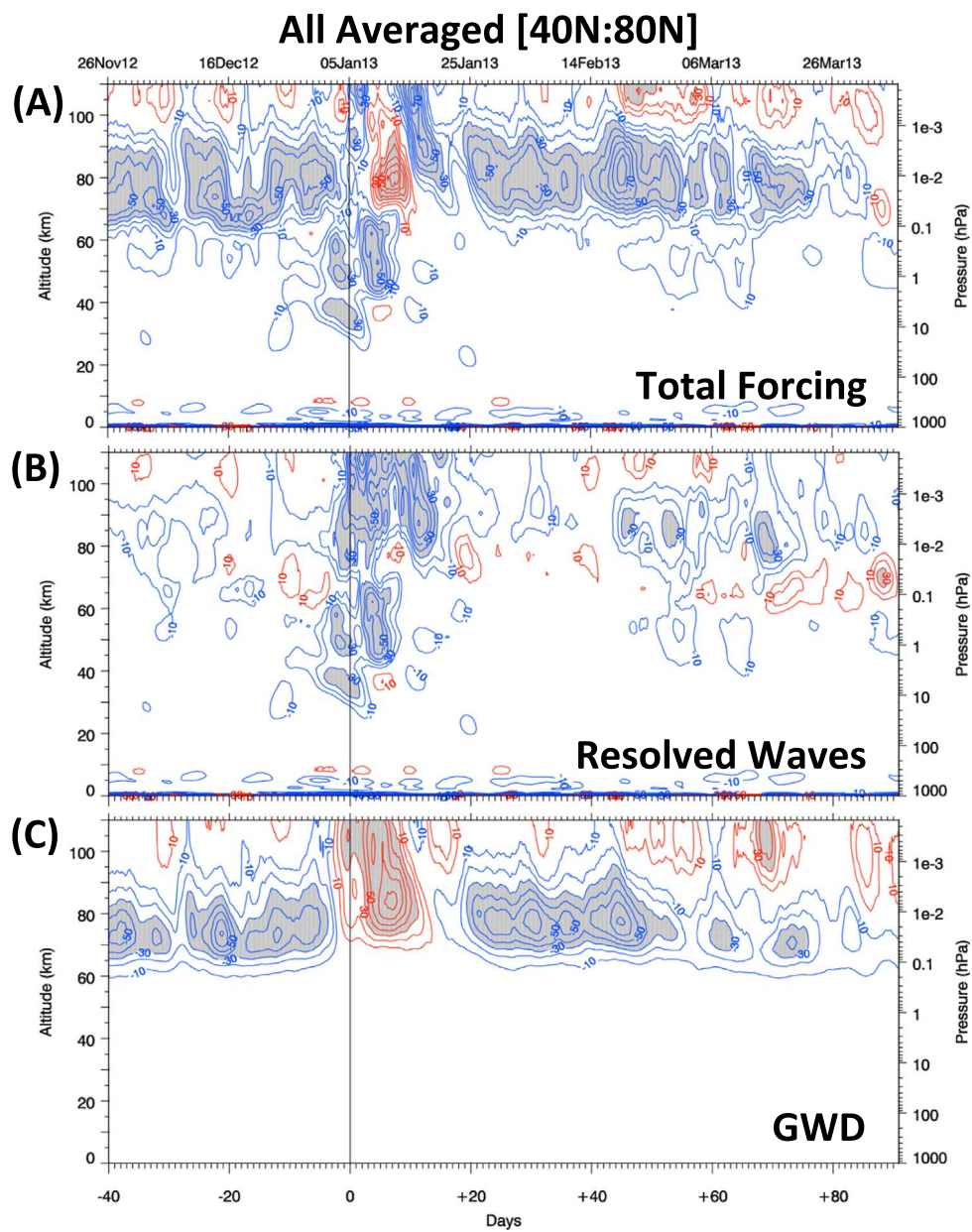

**Fig. 2.** (A) The SD-WACCM total wave forcing, i.e. the sum of resolved waves (EPFD) and parameterized gravity waves (GWD), (B) the resolved wave forcing and (C) parameterized gravity wave forcing. Absolute values greater than 30 m s$^{-1}$ day$^{-1}$ are shaded. All values are averaged from 40°N to 80°N.

Downward motions prevail in the winter polar mesosphere, peaking above the stratopause as expected, except for a short period of about 10 days after the onset when there are ascending motions up to near 95 km. Induced cooling reaching up to the mesopause was also diagnosed in the SD-WACCM composites of a larger number of ESEs by Limpasuvan et al. (2016).

We examine the respective roles of resolved large-scale waves and of parameterized gravity waves in forcing the residual TEM circulation. The time evolution of these two components represented by the EP flux divergence (EPFD) and the gravity wave drag (GWD), as well as their sum, is shown in Fig. 2. Given the broad meridional extent of the forcing, it is averaged over [40–80°N]. The total forcing is westward, driving a poleward and downward mean meridional circulation, and maximizes in the mid and upper mesosphere between 65 and 95 km. The main component forcing this mesospheric summer pole-to-winter pole circulation is the GWD as expected, but there are remarkable changes around the onset, during a period of approximately 20 days. The GWD then turns briefly eastward, hence driving the mesospheric ascent and cooling, before returning to westward. de Wit et al. (2014) showed that the timing of the GWD reversal and the magnitude of the simulated GWD during this event is in broad agreement with the GW forcing derived from radar observations at Trondheim (63.4°N). During a 7-day period between the onset and the stratopause reformation, the westward EPFD is larger above 80 km than the GWD. Since the resolved wave drag results mostly from the longest zonal wavelengths, planetary-scale waves in the MLT region play an important role in the initiation of the elevated stratopause, as first noted by Limpasuvan et al. (2012), and later confirmed by other studies (Chandran et al., 2013; Stray et al., 2015; Limpasuvan et al., 2016).

## 4. Descent from the MLT region: NO

Fig. 3 presents SMR observations of the mean vertical profile of NO averaged in longitude and over [70°N, 90°N], for several days prior to and after the onset of the January 2013 ESE. Also shown is the averaged profile for October and November 2012, used as a pre-winter mean reference profile. Layers of enhanced NO are apparent in the weeks following the ESE onset (days 27–59, for example). The descent of high NO following the January 2013 ESE was shown by Pérot et al. (2014) to be the strongest observed by SMR in the last decade; a similar conclusion was reached using Solar Occultation for Ice Experiment (SOFIE) observations of NO by Bailey et al. (2014). Note that most $NO_x$ are in the form of NO above 70 km, while some conversion between NO and $NO_2$ occurs below.

Fig. 3 also shows the mean polar profile of NO in SD-WACCM (dashed lines, shown for the enhanced diffusion run), the model being sampled as SMR (i.e. with geolocation). It is apparent that there is a low model bias in the pre-warming conditions (e.g., day −25), in the stratosphere and the mesosphere up to $10^{-2}$ hPa. Above that level, i.e. in the MLT reservoir of NO, the profiles are comparable to observations. It is also apparent in Fig. 3 that the descent of high NO in the mesosphere is very weak in SD-WACCM (e.g., days 25, 45, 59). Fig. 4 shows the time evolution throughout the ESE of the polar-averaged NO in the standard and the enhanced diffusion SD-WACCM simulations, as well as their percent-wise relative difference (standard run minus enhanced diffusion run, divided by standard run). The onset begins with strong ascent throughout the mesosphere, making the NO contours deflect upwards. It is followed by a strong descent from the MLT region (e.g., day 10) bringing high NO mixing ratios downward, and the descent then continues throughout the mesosphere (e.g. to day 60). First, one can remark that the enhanced diffusion run has higher NO mixing ratios in the lower and mid mesosphere all throughout the descent. The negative difference fluctuates between −20% and −80%, and the most negative values are found in the NO-rich tongue descending throughout the mesosphere, around day 18 at 75 km. The fluctuating difference arises from the intermittent character of the gravity wave breaking that is triggering the Prandtl number-dependent eddy diffusion.

Enhanced eddy diffusivity has hence clearly increased the lower and mid-mesospheric background levels of NO in SD-WACCM, which had been diagnosed to be too low in case studies of the 2004 or 2009 ESEs (Randall et al., 2015; Siskind et al., 2015). However, there is little difference between the two SD-WACCM simulations above 90 km between the onset and day 20, i.e. when the descent begins. As was shown in Fig. 2, the initial descent from the MLT is forced mostly by the EPFD, which is greater than the GWD. Further inspection of Fig. 4A,B where the total wave forcing for each simulation is overlaid over the NO distribution (white bold contours for negative forcing), clearly reveals the coincidence between the maximum in resolved wave forcing and the initiation of the NO descent.

The relatively weak NO descent from the MLT in SD-WACCM compared to SMR observations is dramatically striking in Fig. 5, which shows the relative difference between SD-WACCM and SMR. The absolute difference is scaled by the mean pre-winter NO profile from SMR to remove the low bias in SD-WACCM NO in the stratosphere and mesosphere in the pre-warming conditions, which is apparent in the profiles of Fig. 3. The noteworthy NO deficit during the descent is nearly identical in both simulations, with or without enhanced eddy diffusion. While the scaled relative difference is on the order of 1 prior and after the ESE event, the strong discrepancy that originates abruptly when the elevated stratopause reforms, is of the order of 20. Thus, it is apparent that the enhanced diffusion obtained by halving the Prandtl number does not completely eliminate the negative NO bias during the descent.

To further diagnose how the average descent of NO compares between the model and the SMR observations, we infer the descent rate from the NO isopleths, using the method applied to MLS measurements by Lee et al. (2011) and to SOFIE measurements by Bailey et al. (2014). Given a choice of NO mixing ratio values, and ensuring a measurement response greater than 0.67 and a signal-to-noise ratio greater than 10, the profile of the descent using SMR data is retrieved between $10^{-3}$ and 0.07 hPa at the most. Altitudes of polar-averaged [70–90°N] NO mixing ratios were tracked over two periods, before and after the onset (days −100 to −15, incorporating 12 measurement days, and days 5–35, incorporating 5 measurement days, respectively). A linear fit to those altitudes over the span of each period was drawn to infer a descent rate. The same procedure was applied to SD-WACCM mixing ratios over the same height range as SMR, and the results are shown on Fig. 6. While comparable prior to the onset, the descent rate in SD-WACCM after the onset diverges from the SMR rate around 0.002 hPa, becomes increasingly slower, the ratio reaching a factor of 1/2 near 0.08 hPa.

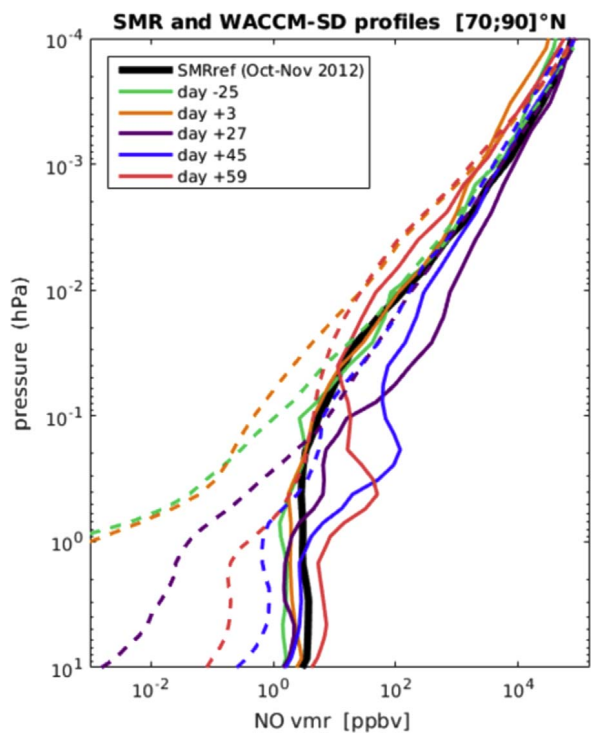

**Fig. 3.** Vertical profile of NO averaged over [70–90°N] measured by SMR (solid lines) and modelled by SD-WACCM (dashed lines) during the winter 2012/13. Day numbers are relative to the stratospheric warming onset date. Also shown (thick black line) is the SMR profile averaged over October and November 2012, used as a pre-warming reference profile. SD-WACCM is sampled as SMR.

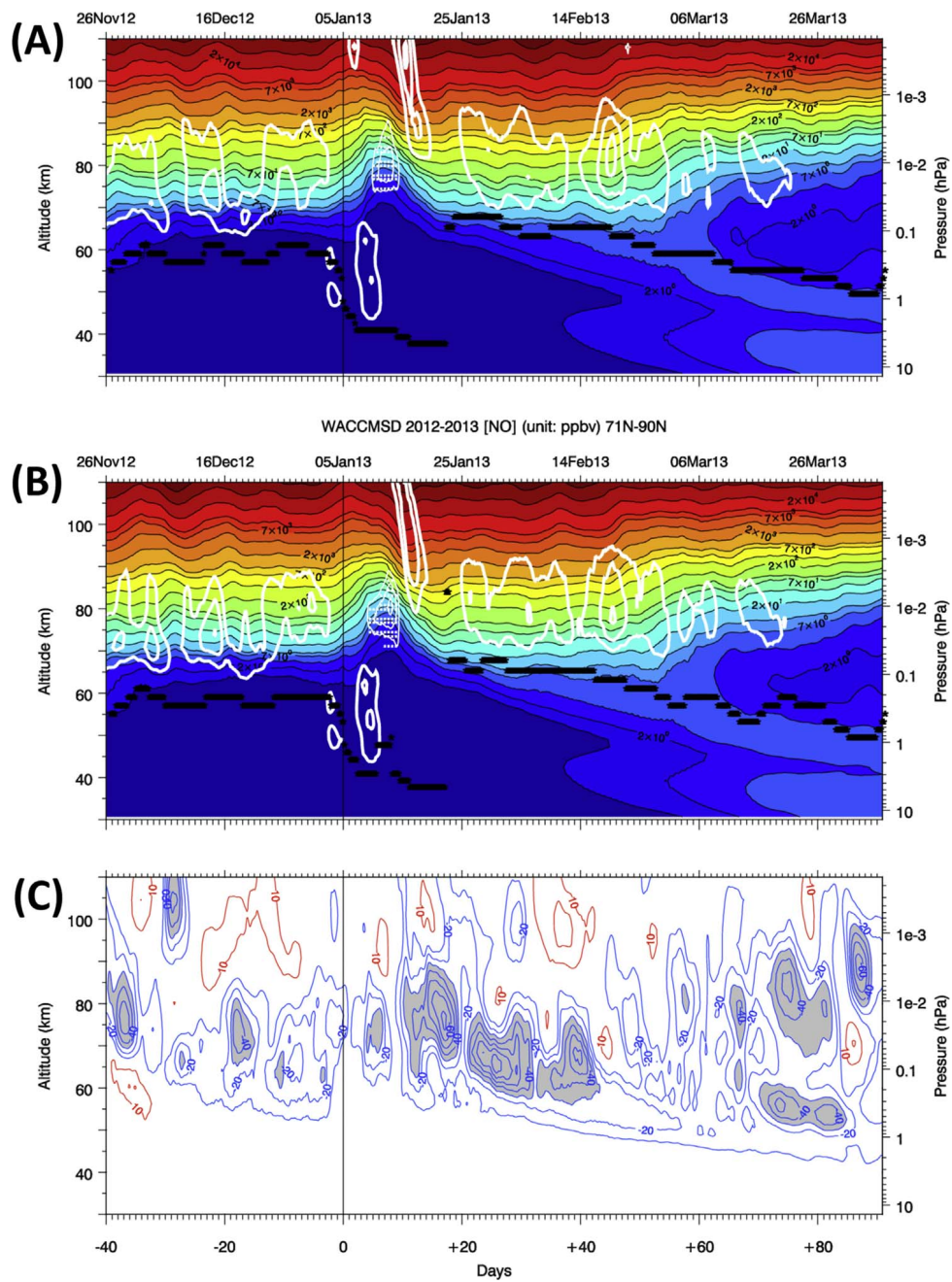

**Fig. 4.** Averaged SD-WACCM mixing ratio of NO over [70°N–90°N] in the two simulations with Prandtl number 4 (A), and (enhanced diffusion) with Prandtl number 2 (B) and (C) their relative difference in %. Also shown is the total wave forcing with absolute values greater than 40 m s$^{-1}$ day$^{-1}$ (white contours), as in Fig. 2A. Positive (eastward) forcing values are stippled and shown by less thick white contours. Shaded areas correspond to relative differences more negative than −30%.

In Fig. 5, it appears that the NO deficit at the initiation of the descent arises first near 6 $10^{-4}$ hPa. In their study of the 2009 ESE with the HAMMONIA (Hamburg Model of Neutral and Ionized Atmosphere) model, Meraner and Schmidt (2016) also found that the downward transport across the mesopause following the onset was largely driven by vertical advection, while eddy diffusion had limited impact in the upper mesosphere. In accordance with our results, the descent was larger at the time of the stratopause reformation (their Fig. 6). Furthermore, doubling the eddy diffusion did not bring their model polar-averaged profile of NO much closer to the Michelson Interferometer for Passive Atmospheric Sounding (MIPAS) observations (their Fig. 3). The monthly-averaged descent rate following the onset, as inferred from SMR NO (Fig. 6, blue line), which is slightly below 10 mm/s near 0.02 hPa, is nearly identical to the corresponding mean descent rate deduced from the evolution of idealised tracers in HAMMONIA (their Fig. 4h). Meraner and Schmidt (2016) also noted that the transport is enhanced if 6-hourly rather than daily averaged TEM vertical velocities are used. In our case, we used 3-hourly values.

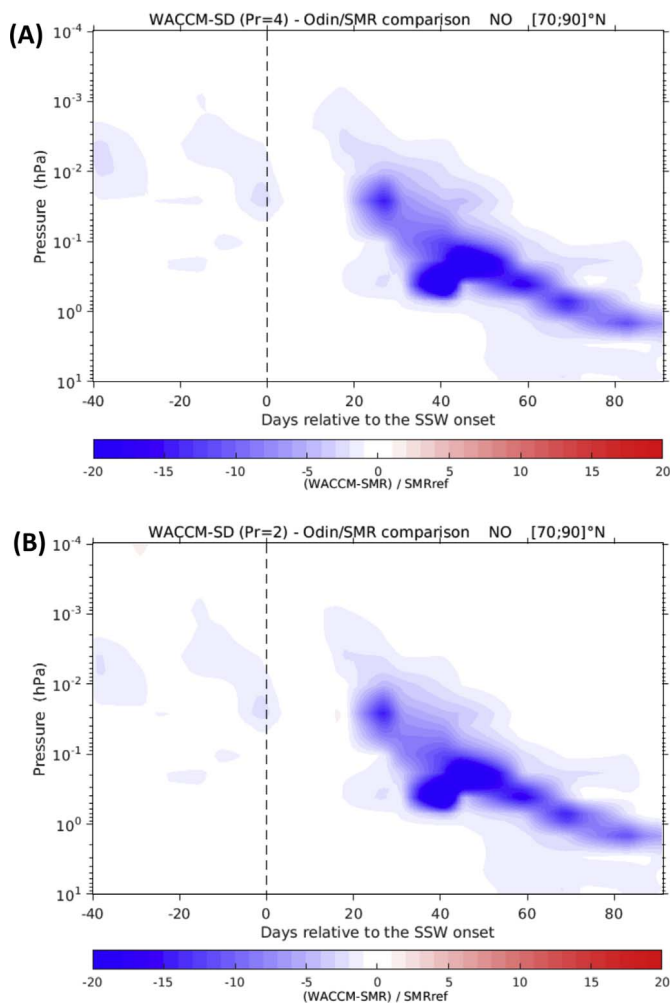

**Fig. 5.** Relative difference between the averaged NO over [70°N–90°N] between SD-WACCM and SMR, scaled by the mean pre-warming reference profile (as in Fig. 3), for the two simulations with Prandtl number 4 (a), and Prandtl number 2 (b). SD-WACCM is sampled as SMR.

It is worth noting that the inferred descent from the tracer isopleths is different from the vertical TEM velocity (w*) (also shown in Fig. 6), since the former is also affected by meridional mixing and vertical diffusion, which both depend on the background NO gradient. Meridional mixing is important. As the vortex re-establishes itself following the onset, poleward transport of less abundant NO from lower latitudes provides a "dilution effect", flattening the isopleths and making the inferred descent rate smaller than w*.

## 5. Descent from the MLT region: $H_2O$

Water vapour has been widely used to infer descent of dry mesospheric air during ESEs (Orsolini et al., 2010; Bailey et al., 2014). The strong descent throughout the mesosphere following the ESE in January 2013 diagnosed in NO, also left a strong imprint in the mesospheric water vapour distribution, as revealed by the SMR observations. In the winter polar region, the climatological poleward and downward mean meridional circulation transitions near 90 km to an opposite cell aloft, characterised by polar ascent (e.g. Lossow et al., 2009; Smith et al., 2011). Since the $H_2O$ vertical gradient is negative, this uplift leads to a distinctive climatological dome of $H_2O$-enriched air over the pole in winter, extending upwards from approximately 90 km. This annual climatological cycle is shown in Fig. 7A which is updated from Lossow et al. (2009) and covers years 2007–2015. Inspection of the $H_2O$ mixing ratio evolution in Fig. 7B (e.g. the 0.05 ppmv isopleth) reveals that, immediately following the onset of the ESE, the descent of $H_2O$-poor air from the MLT region clearly dents that dome of $H_2O$-enriched air. Note that, as we emphasize different features in the climatological distribution in Fig. 7A than in the distribution around the onset in Fig. 7B, the axes and the contouring are slightly different. The key point is that the $H_2O$ observations clearly reveal that the mesospheric descent initiates in the MLT shortly following the onset. The descent may be associated with the warmer mesopause that reforms near 100 km following its prior downward shift (Fig. 1B, after day 12) noted by Kishore et al. (2016). The descent of dry air then continues throughout the mesosphere and into the upper stratosphere in the following two months (Fig. 7B, e.g., the 2.5 ppmv contour). In SD-WACCM, the diagnosis of NO clearly showed that descent from the MLT region is too weak. Consequently, $H_2O$ mixing ratios are higher than observed by SMR (not shown for brevity).

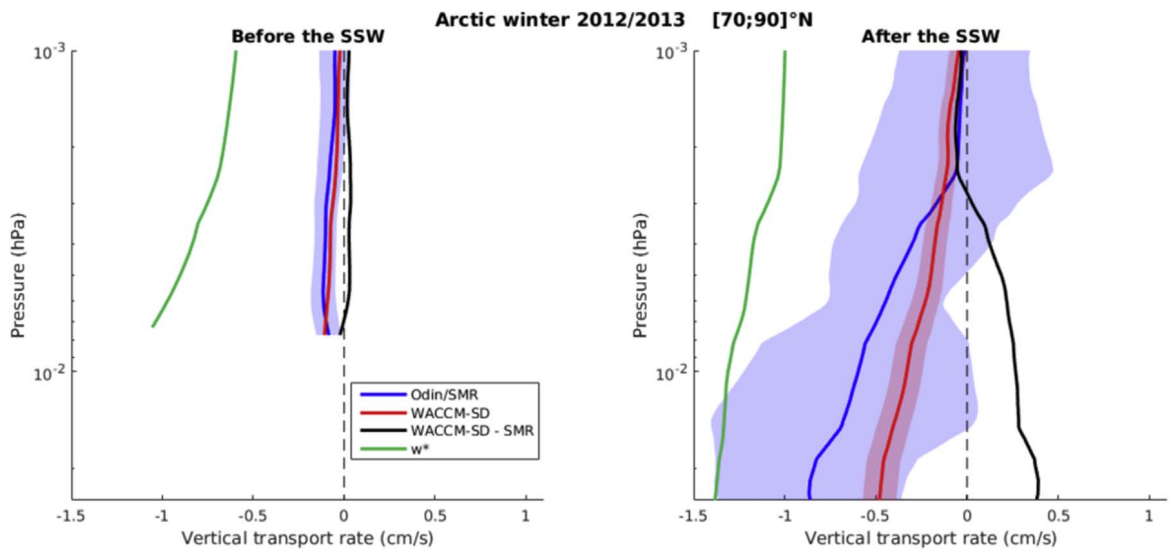

**Fig. 6.** Polar-averaged [70–90°N] inferred descent rate from the NO isopleths in SMR and SD-WACCM and their difference (SD-WACCM minus SMR) before and after the onset (days −100 to −15, incorporating 12 measurement days, and days 5–35, incorporating 5 measurement days, respectively). The SD-WACCM TEM vertical velocity (w*) is also plotted. (For interpretation of the references to color in this figure, the reader is referred to the web version of this article.)

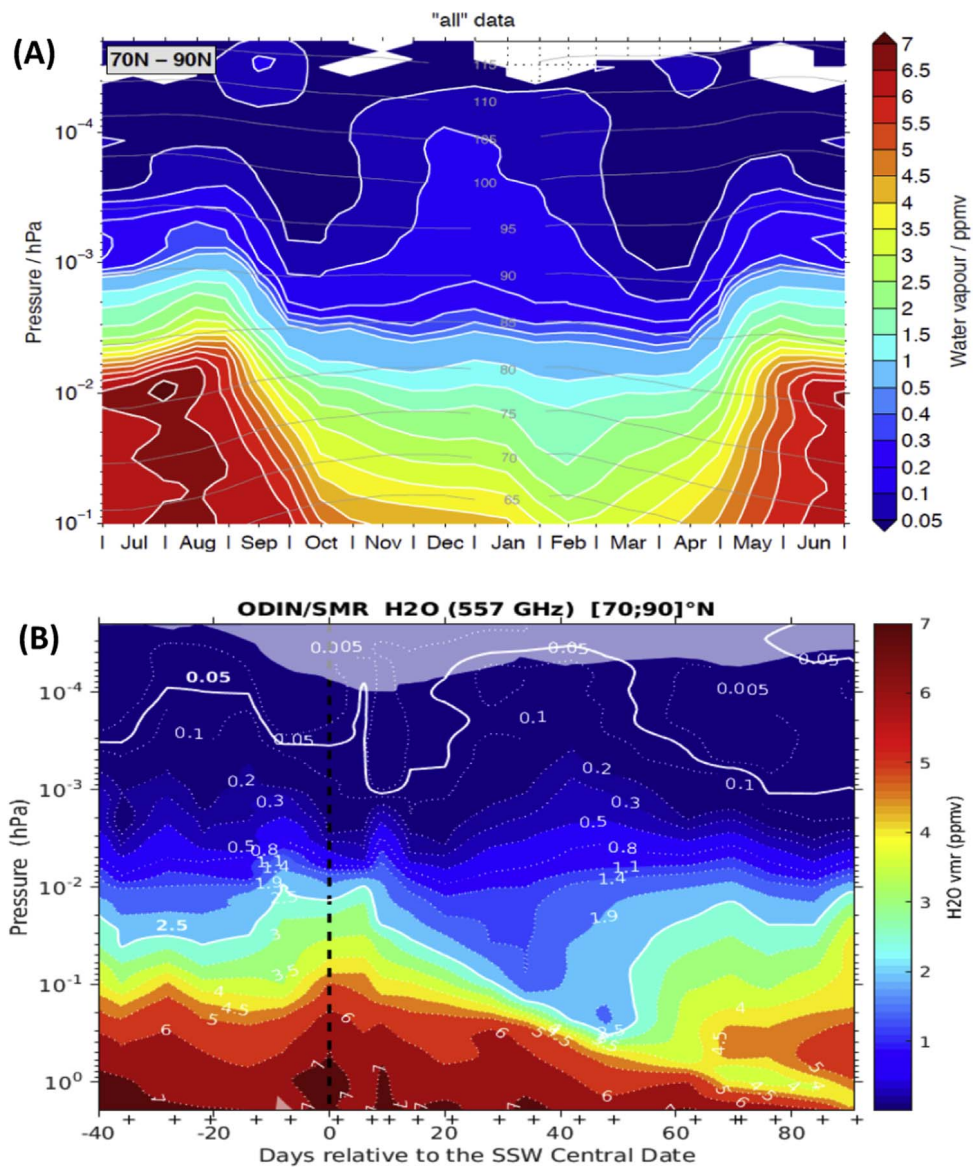

**Fig. 7.** (A) Climatological annual cycle of SMR water vapour mixing ratio (in ppmv), averaged over [70–90°N], from 2007 to 2015. Approximate altitude range are also indicated (thin blue contours). (B) Time-Pressure cross-section of SMR water vapour mixing ratio (in ppmv), averaged over [70°N, 90°N]. Mixing ratio of 0.05 and 2.5 ppmv are highlighted as thick white contours. Markings (+) along x-axis indicate days when measurements were obtained. The shaded or white areas above $10^{-4}$ hPa indicates a low measurement response. Note the slightly different vertical range and contouring between (A) and (B).

## 6. Enhancement of high-latitude tides

Another characteristic of the ESEs is the excitation of the SW2 tide, as mentioned in the Introduction. Meteor radar observations of the semidiurnal zonal and meridional tidal winds over Trondheim (63.4°N) during January 2013 are shown in Fig. 8C,D between 82 and 97 km. They reveal a clear enhancement of the semidiurnal tide with a peak around 14 days after the onset, albeit there is a small difference in the timing of the maximum between the two wind components. The zonal and meridional winds associated to the migrating semidiurnal (SW2) tide in SD-WACCM are shown in Fig. 8A,B at 65°N, but with a different altitude range, from 80 to 120 km. This range is chosen to show more completely the vertical amplification of the model tides, since they are generally too weak, as noted in earlier studies (e.g. Smith, 2012). Specifically, the magnitude of the semidiurnal tide observed by the radar and its enhancement is not reproduced in SD-WACCM below 97 km, and only higher up is there a remarkable agreement between the model and the radar observations with regard to the timings of the tidal wind maxima between January 20 and 25.

A single-station approach such as this will measure the vector sum of all the Hough modes of both the migrating and non-migrating components of the semidiurnal tide at the given latitude and longitude of the station. While previous work (e.g. Iimura et al., 2010) has shown the evidence of non-migrating components to the semidiurnal tide in the high latitude northern hemisphere MLT region, the migrating (SW2) component is expected to be the dominant component at these latitudes in winter (e.g. Manson et al., 2009); its relative behaviour is therefore directly comparable to the model. There is an abundant literature on the tidal local effects on trace constituent abundances (e.g. Smith, 2012 and references therein). We are interested here in how the EP flux associated to the SW2 tide, which maximizes in northern subtropics, could contribute to the forcing of the residual TEM circulation into high latitudes. Fig. 9 shows the resolved, gravity and total wave forcings averaged over the latitude band [40°N, 80°N] and between 90 km and 130 km. Note that the pre-warming gravity wave drag is eastward in that altitude range (as opposed to westward below 90 km, in Fig. 2). Fig. 9 also displays the small part of the resolved wave forcing that arises from the SW2 tide (blue line) and from the diurnal migrating tide (DW1, red line), after having been multiplied by a factor of 5 for clarity. It appears that not only are the tidal forcings a small contribution to the resolved wave forcing, but also that the SW2 tidal enhancement lags the peak in resolved wave forcing and the period of strong descent. In their model case study of the 2009 ESE using WACCM-X, Sassi et al. (2013) found that the DW1 tide was also enhanced following the onset, reaching comparable amplitude to SW2 tide in terms of temperature at 100 km. During the 2013 event, DW1 has smaller zonal wind amplitude than SW2, by about a factor 5, both in SD-WACCM and the radar data (not shown).

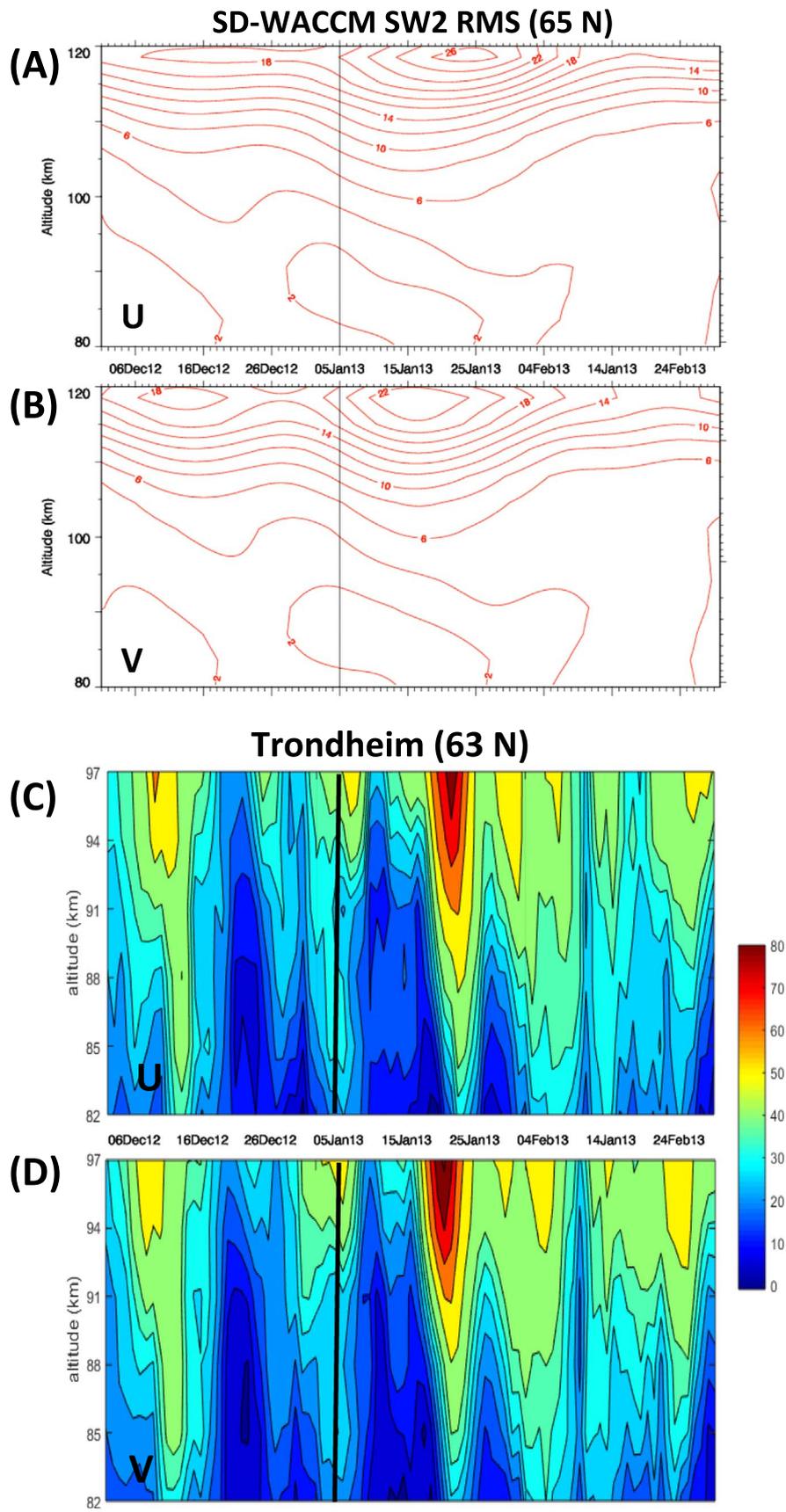

**Fig. 8.** Semidiurnal migrating tide root-mean-square amplitude from SD-WACCM at 65°N in zonal (A) and meridional (B) wind, and corresponding radar observations of the semidiurnal tide at Trondheim at 63 N (C, D). Units are m s$^{-1}$. Note the different height range and contouring between (C, D) and (A,B).

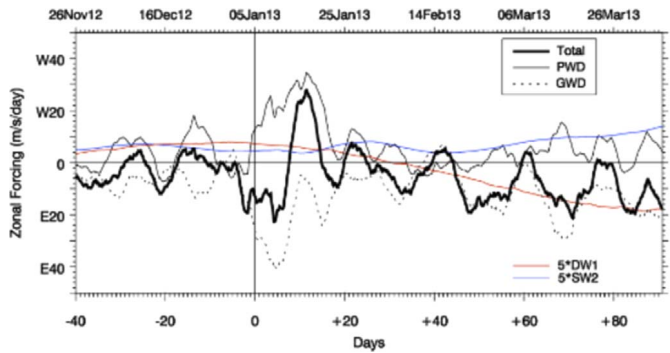

**Fig. 9.** SD-WACCM total as well as resolved wave (PWD) and gravity wave (GWD) forcings averaged over [40–80°N] and between 90 km and 130 km. Westward and eastward forcings are respectively labelled with W or E on the y-axis. Component of that forcing from migrating diurnal tide (red) and from migrating semidiurnal tide (blue) are also shown, with their values are multiplied by a factor of 5. (For interpretation of the references to color in this figure legend, the reader is referred to the web version of this article.)

## 7. Discussion and conclusions

Using simulations with WACCM nudged by meteorological analyses, global satellite observations, and local radar observations, this study examines the recent ESE that occurred in January 2013, and its impact on high NO and dry air descent into the stratosphere.

Using the same model, Limpasuvan et al. (2016) considered a composite of 13 ESEs, largely dominated by displacement events over split events. The resolved wave forcing initiating the MLT descent and the tidal amplification of SW2 *following* the onset were both featured in the composite. Hence, we have no reason to believe that the under-estimation of NO would be linked to the fact that the 2013 event is a vortex split. However, as discussed in that paper (Section 7), the vortex-splitting event of 2009 was associated with mesospheric *precursory* planetary wave-2 activity (e.g., Coy et al., 2011). The latter might further interact with the tidal components. In our case, the EPFD remains very weak in the weeks prior to the onset in the 80–100 km altitude range. The role of mesospheric precursor waves in vortex splitting events deserves further attention.

The descent of NO is under-estimated in SD-WACCM, despite enhancing the eddy diffusion (via the equivalent Prandtl number) by a factor 2. The descent initiates in the MLT shortly following the onset, and we have shown that the resolved (mostly planetary) waves play a crucial role in forcing the initial downward TEM circulation. We further note that the planetary waves generated in SD-WACCM in the MLT region are comparable to observations in their timing and zonal wavenumber. Fig. 10 shows a Høvmuller plot of temperature at 52°N and 91 km for SABER and for SD-WACCM, clearly revealing a westward-propagating wave-1 after the onset. Note that this is a raw comparison and SABER weighting function (kernel) was not applied to the model results. The peak amplitude of this wave in SD-WACCM (about 35 K) is also comparable to the one observed by MLS at 63°N and 0.01 hPa during this event in de Wit et al. (2015a), their Fig. 5). Yet, the resolved waves do not seem to generate a large enough TEM downward motion to generate a sufficient downward transport of NO to match satellite observations, despite the model having high enough background abundance of NO above 0.001 hPa. This may indicate issues with planetary wave transience, dissipation or fluxes. Another possibility is that the gravity wave breaking in the model around the onset should also extend into the upper mesosphere, hence extending the region of westward GWD (see Fig. 2) well above 90 km.

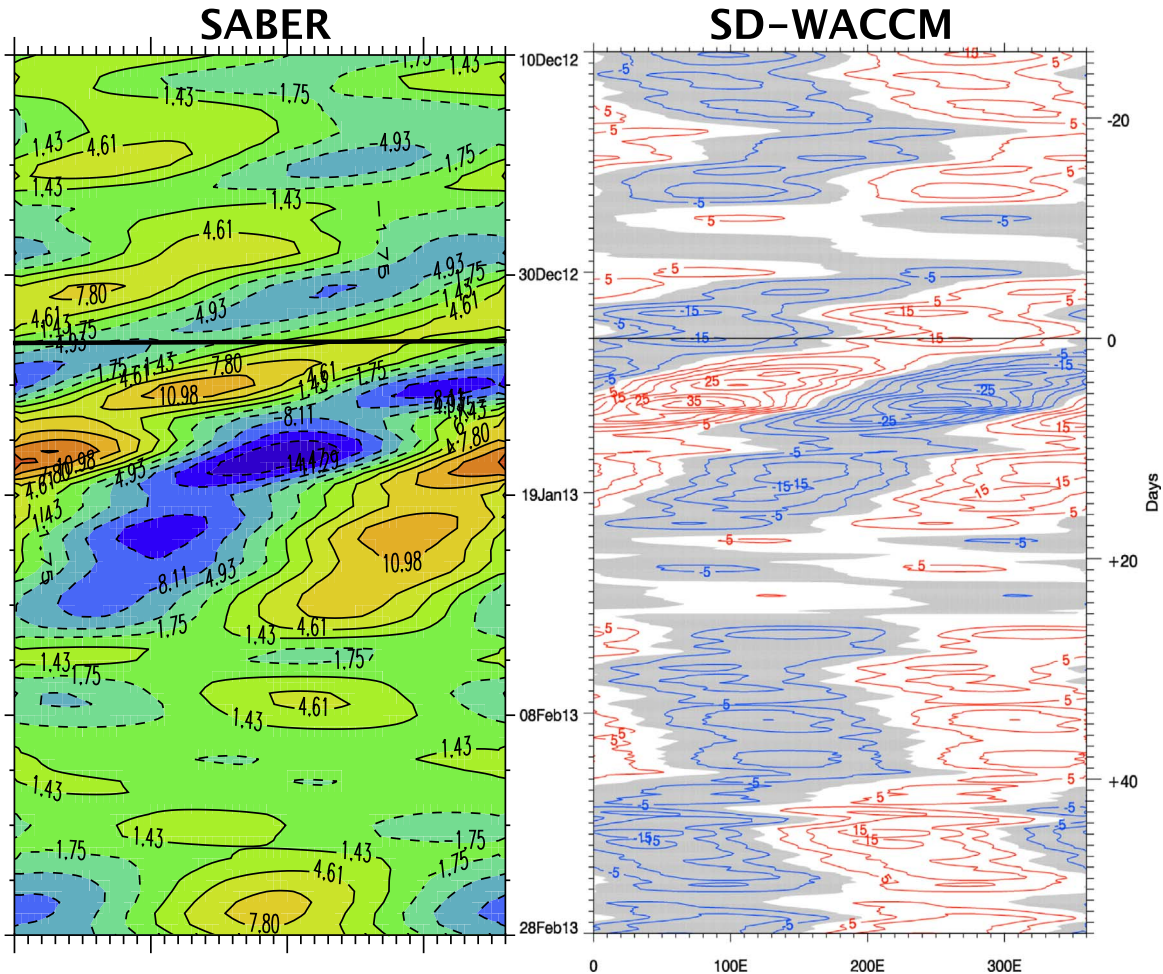

**Fig. 10.** Høvmuller diagram for the zonal wavenumber 1 temperature perturbations (°K) at 91 km and 52°N for SABER and for SD-WACCM. Shaded area in grey corresponds to negative values. The horizontal line denotes the onset.

It clearly appears from the comparison with meteor radar winds that the tides are too weak in SD-WACCM as noted in previous studies (e.g. Smith, 2012). Addressing the multiple causes for the weak tidal amplitudes in the model is beyond the scope of this paper, but excessive damping in the MERRA re-analyses or the model, insufficient vertical resolution in the middle atmosphere, damping by the gravity wave parametrisation, and incorrect tidal waveguide representation could be invoked. We surmise that if the amplitude of the WACCM tides at 90 km were around 80 m s$^{-1}$, as observed, then the tidal contribution to the westward forcing in the lower thermosphere would be much greater. While a stronger tidal forcing could possibly augment the downward TEM circulation, it nevertheless appears from Fig. 9 that the tidal amplification lags the high-altitude initiation of the descent, and that planetary wave forcing appears to be the key component for the brief strengthening of the residual circulation at the formation of the elevated stratopause.

## Acknowledgements

YO is supported by the Research Council of Norway (grant HEPPA-Norway #222390). YO, RH and PE are partially supported at the Birkeland Centre for Space Science by the Research Council of Norway/CoE under contract 223252/F50. V.L. is supported in part by the Large-Scale Dynamics Program at the National Science Foundation (NSF) under award AGS-1116123 and AGS-1624068, and the Kerns Palmetto Professorship supported by the Coastal Carolina University's Provost office. We also thank the support of the International Space Science Institute (ISSI) provided within the study group on the added-value of chemical data assimilation in the stratosphere and upper-troposphere. The authors are grateful to Doug Kinnison, Dan Marsh and Rolando Garcia at NCAR for help with enhanced eddy diffusion simulations.